\documentclass[11pt,fleqn]{article}

\usepackage{comment}

\usepackage{amsmath,amssymb,graphicx,mdframed,multicol,pslatex,algorithm2e,verbatim,wrapfig}

\includecomment{series}
\excludecomment{notseries}
\includecomment{book}

\newif\ifindexmargin \indexmargintrue
\def\marginindex#1{}
\newcommand{\indexterm}[1]{\emph{#1}\index{#1}%
    \marginindex{#1}}

\makeatletter
\usepackage{acronym}
\newcommand\indexac[1]{\emph{\ac{#1}}%
  \edef\tmp{\noexpand\index{%
    \expandafter\expandafter\expandafter
        \@secondoftwo\csname fn@#1\endcsname%
    @\acl{#1} (#1)}}\tmp}
\newcommand\indexacp[1]{\emph{\ac{#1}}%
  \edef\tmp{\noexpand\index{%
    \expandafter\expandafter\expandafter
        \@secondoftwo\csname fn@#1\endcsname%
    @\acl{#1} (#1)}}\tmp}
\newcommand\indexacf[1]{\emph{\acf{#1}}%
  \edef\tmp{\noexpand\index{%
    \expandafter\expandafter\expandafter
        \@secondoftwo\csname fn@#1\endcsname
    @\acl{#1} (#1)}}\tmp}
\newcommand\indexacs[1]{\emph{\acs{#1}}%
  \edef\tmp{\noexpand\index{%
    \expandafter\expandafter\expandafter
        \@secondoftwo\csname fn@#1\endcsname
    @\acl{#1} (#1)}}\tmp}
\newcommand\indexacstart[1]{%
  \edef\tmp{\noexpand\index{%
    \expandafter\expandafter\expandafter
        \@secondoftwo\csname fn@#1\endcsname
    @\acl{#1} (#1)|(}}\tmp}
\newcommand\indexacstartbf[1]{%
  \edef\tmp{\noexpand\index{%
    \expandafter\expandafter\expandafter
        \@secondoftwo\csname fn@#1\endcsname
    @\acl{#1} (#1)|(textbf}}\tmp}
\newcommand\indexacend[1]{%
  \edef\tmp{\noexpand\index{%
    \expandafter\expandafter\expandafter
        \@secondoftwo\csname fn@#1\endcsname
    @\acl{#1} (#1)|)}}\tmp}
\newcommand\indexacsub[2]{\emph{\ac{#1}}%
  \edef\tmp{\noexpand\index{%
    \expandafter\expandafter\expandafter
        \@secondoftwo\csname fn@#1\endcsname%
    @\acl{#1} (#1), #2}!#2}\tmp}
\makeatother

\acrodef{API}{Application Programmer Interface}
\acrodef{BFS}{Breadth-First Search}
\acrodef{BH}{Barnes-Hut}
\acrodef{BLAS}{Basic Linear Algebra Subprograms}
\acrodef{BSP}{Bulk Synchronous Parallelism}
\acrodef{CAF}{Co-Array Fortran}
\acrodef{CG}{Conjugate Gradients}
\acrodef{CREW}{Concurrent Read, Exclusive Write}
\acrodef{CSP}{Communication Sequential Processes}
\acrodef{DAG}{Directed Acyclic Graph}
\acrodef{DSL}{Domain-Specific Language}
\acrodef{DTD}{Distributed Termination Detection}
\acrodef{FDM}{Finite Difference Method}
\acrodef{FE}{Finite Element}
\acrodef{FEM}{Finite Element Method}
\acrodef{FMM}{Fast Multipole Method}
\acrodef{FSM}{Finite State Machine}
\acrodef{GA}{Global Arrays}
\acrodef{GPU}{Graphics Processing Unit}
\acrodef{GS}{Gauss-Seidel}
\acrodef{HPC}{High Performance Computing}
\acrodef{HPF}{High Performance Fortran}
\acrodef{IMP}{Integrative Model for Parallelism}
\acrodef{IR}{Intermediate Representation}
\acrodef{ISA}{Incremental Single Assignment}
\acrodef{KNL}{Knights Landing}
\acrodef{MD}{Molecular Dynamics}
\acrodef{MPI}{Message Passing Interface}
\acrodef{NUMA}{Non-Uniform Memory Access}
\acrodef{OO}{Object-Oriented}
\acrodef{PDE}{Partial Differential Equation}
\acrodef{PGAS}{Partitioned Global Address Space}
\acrodef{PRAM}{Parallel Random Access Machine}
\acrodef{RMA}{Remote Memory Access}
\acrodef{RPC}{Remote Procedure Call}
\acrodef{SIMD}{Single Instruction Multiple Data}
\acrodef{SIMT}{Single Instruction Multiple Thread}
\acrodef{SA}{Simulated Annealing}
\acrodef{SMP}{Symmetric Multi-Processor}
\acrodef{SSA}{Static Single Assignment}
\acrodef{SPMD}{Single Program Multiple Data}
\acrodef{SSSP}{Single-Source Shortest Path}
\acrodef{UPC}{Universal Parallel C}

\usepackage{etoolbox}

\usepackage{xr-hyper}
\begin{book}
\usepackage[pdftex,colorlinks]{hyperref}
\end{book}

\hypersetup{bookmarksopen=true}
\def\heading#1{\paragraph*{\textbf{#1}\kern1em}\ignorespaces}

\expandafter\ifx\csname corollary\endcsname\relax
    
\fi
\expandafter\ifx\csname lemma\endcsname\relax
    
\fi
\expandafter\ifx\csname theorem\endcsname\relax
    \newtheorem{theorem}{Theorem}
\fi
\expandafter\ifx\csname proof\endcsname\relax
 
\fi
\expandafter\ifx\csname definition\endcsname\relax
  
\fi
\expandafter\ifx\csname example\endcsname\relax
  
\fi
\expandafter\ifx\csname remark\endcsname\relax
  
\fi

\def\pred{\mathop{\mathit{pred}}}

\usepackage{framed}

\usepackage{mdframed}

\def\n{\bgroup\tt\catcode`\_=12 \let\next=}

\def\twocode {\afterassignment\twocodeb\def\nexta}
\def\twocodeb{\bgroup \catcode`\_=12\relax
              \afterassignment\twocodec\global\def\nextb}
\def\twocodec{\egroup 
  \par\smallskip
  \hskip\unitindent $\vcenter{\hsize=.37\hsize$\nexta$}\quad
   \vcenter{\hsize=.5\hsize\footnotesize\tt\nextb}$
  \par\smallskip
}

\excludecomment{proposal}
\includecomment{longstory}
\includecomment{lrsx}
\includecomment{dataflow}
\excludecomment{public}
\includecomment{implementation}
\includecomment{paper}

\begin{proposal}
\excludecomment{lrsx}
\excludecomment{paper}
\end{proposal}

\includecomment{insulting}
\includecomment{fullmonty}
\begin{public}
\excludecomment{fullmonty}
\excludecomment{insulting}
\excludecomment{impquestion}
\excludecomment{inprogress}
\end{public}

\begin{fullmonty}

\newbox\qbox

\end{fullmonty}

\includecomment{nonlrsx}
\begin{lrsx}
\excludecomment{nonlrsx}
\end{lrsx}

\includecomment{shortstory}
\begin{longstory}
\excludecomment{shortstory}
\end{longstory}
\includecomment{theory}
\begin{shortstory}
\excludecomment{theory}
\end{shortstory}

\includecomment{lrsx}
\includecomment{mvpexample}
\includecomment{bh}
\includecomment{programming}
\includecomment{partialdist}
\includecomment{dataflow}

\usepackage{outliner}
\OutlineLevelStart0{\section{#1}}
\OutlineLevelStart1{\subsection{#1}}
\OutlineLevelStart2{\subsubsection{#1}}

\def\n{\bgroup\tt\catcode`\_=12 \let\next=}

\title{Task Graph Transformations for Latency Tolerance}
\author{Victor Eijkhout\thanks{{\tt
      eijkhout@tacc.utexas.edu}, Texas Advanced Computing Center, The
    University of Texas at Austin}}

\begin{document}
\maketitle

\begin{abstract}
The Integrative Model for Parallelism (IMP) derives a task graph from
a higher level description of parallel algorithms.
In this note we show how task graph transformations can be used to
achieve latency tolerance in the program execution. We give a formal
derivation of the graph transformation, and show through simulation
how latency tolerant algorithms can be faster than the naive execution
in a strong scaling scenario.
\end{abstract}

\section{Motivation}

On clusters the cost of communication can be high relatively to
the cost of computation. Hence, the \indexterm{overlapping computation
  and communication} (also known as \indexterm{latency hiding}) has
long been a goal of parallel programming.
On shared memory processors
with explicitly managed \indexterm{scratchpad} memory there is an
equivalent phenomenon: if data can be pushed to the scratchpad well in
advance of it being needed, we now hide the memory, rather than
network, latency.

Various techniques for latency hiding have been used. For instance,
the PETSc library~\cite{GrSm:petsc} splits the matrix-vector product
in local and non-local parts, so that the former can overlap the
communication of the latter.
Related, redundant computation in order to avoid communication is an
old idea~\cite{OpJo:improved-ssor}.

In this report we will show that latency hiding, and general latency
tolerance, can be achieved by task graph transformations. For this we
use the formalism of the \ac{IMP}, previously defined
in~\cite{Eijkhout:mathematical2016arxiv}.

There is considerable work in the context of iterative methods for
linear system to mitigate the influence of communication.
\begin{itemize}
\item Reformulation of CG-like methods to reduce the number of inner
  products~\cite{ChGe:sstep,DAzEijRo:ppscicomp,Sa:practicalKrylov,Me:multicg}.
\item Multi-step methods that combine inner products, and can have
  better locality properties~\cite{ChGe:sstep}.
\item
  Overlapping either the preconditioner
  application or the matrix-vector product with a
  collective\cite{dehevo92:acta}. We will give a new variant, based
  on~\cite{Gropp:libraries}, that overlaps both.
\end{itemize}
Recently, the notion of redundant computation was revisited by Demmel
\emph{et al.}~\cite{Demmel2008IEEE:avoiding}, in so-called
`communication avoiding' methods. We will show how such methods
naturally arise in the IMP framework. This will be the main result of
this note.

\section{Communication avoiding}

We explain the basic idea of the `communication avoiding' scheme. This
was originally proposed for iterative methods, such as $s$-step CG; in
the next section we will show that the IMP framework can realize this
in general.


In \ac{PDE} methods, a repeated sequence of sparse matrix-vector products
is a regular occurrence.
Typically, the sparse matrix can best be viewed as an operator
on a grid of unknowns, where a new value is some combination of values
of neighbouring unknowns.
In a parallel context this means that in order to evaluate the matrix-vector
product $y\leftarrow Ax$ on a processor, that processor needs to obtain the $x$-values
of its \indexterm{ghost region}. Under reasonable assumptions on the partitioning
of the domain over the processors, the number of messages involved will be fairly
small:
in a \ac{FEM} or \ac{FDM} context,
the number of messages is $O(1)$ as~$h\downarrow\nobreak 0$.

Since there is little data reuse, and in the sparse case not even 
spatial locality, it is normally concluded that the sparse
product is largely a \emph{bandwidth-bound algorithm}. 
Looking at just a
single product there is not much we can do about that. 
However, 
if a number of such products is performed in a row, for instance as the steps
in a time-dependent process, there may be rearrangements
of the operations that lessen the bandwidth demands, typically by lessening the
latency cost.

Consider as a simple example
\begin{equation}
\forall_i\colon x^{(n+1)}_i = f\bigl( x^{(n)}_i, x^{(n)}_{i-1}, x^{(n)}_{i+1} \bigr)
\label{eq:3p-average}
\end{equation}
and let's assume that the set $\{x^{(n)}_i\}_i$ is too large to fit 
in cache.
This is a model for, for instance, the explicit scheme for the heat
equation in one space dimension.
In the ordinary computation, where we first compute all~$x^{(n+1)}_i$, 
then all~$x^{(n+2)}_i$, the intermediate values at level~$n+1$
will be flushed from the cache
after they were generated, and then brought back into cache as input for the
level $n+2$ quantities.

However,
if we compute not one, but two iterations, the intermediate values
may stay in cache.
Consider $x^{(n+2)}_0$: it requires $x^{(n+1)}_0,x^{(n+1)}_1$,
which in turn require $x^{(n)}_0,\ldots,x^{(n)}_2$.

\begin{figure}[ht]
\includegraphics[scale=.1]{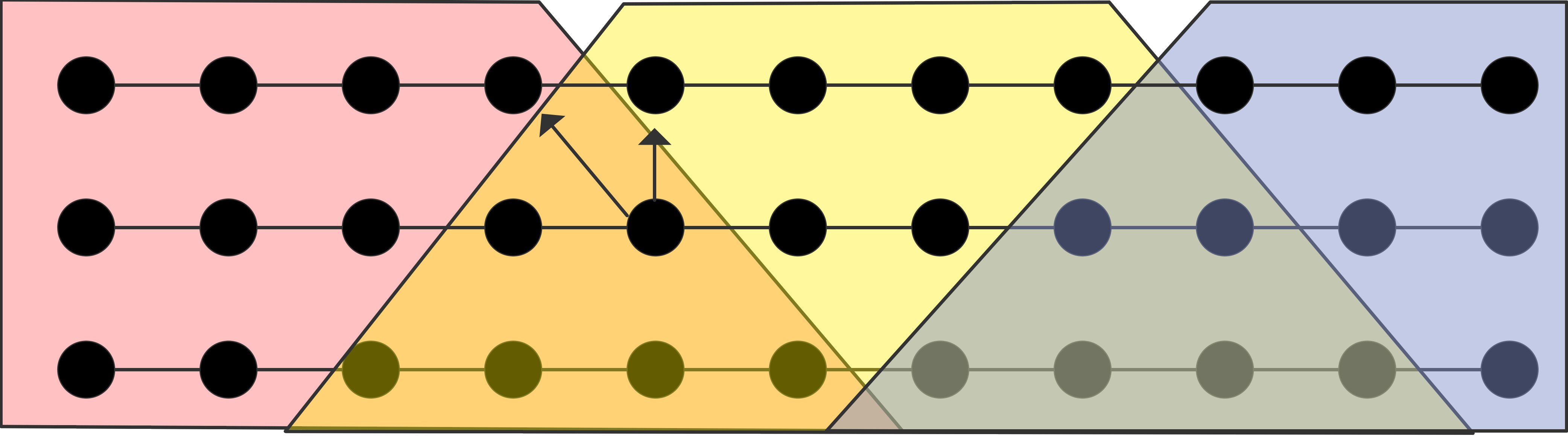}
\caption{Computation of blocks of grid points over multiple iterations}
\label{fig:grid-update-overlap}
\end{figure}

Now suppose that we are not interested in the intermediate results, but
only the final iteration. Figure~\ref{fig:grid-update-overlap} shows
a simple example.
The first processor computes 4~points on level $n+2$. For this it needs 5~points
from level $n+1$, and these need to be computed too, from 6~points on level~$n$.
We see that a processor apparently needs to collect a \indexterm{ghost region}
of width two, as opposed to just one for the regular single step update.
One of the points computed by the first processor is $x^{(n+2)}_3$,
which needs $x^{(n+1)}_4$. This point is also needed for the computation
of $x^{(n+2)}_4$, which belongs to the second processor.

The easiest solution is to let this sort of point on the intermediate
level \emph{redundantly computed}\index{redundant computation}, in 
the computation of both blocks where it is needed, on two different processors.

\begin{itemize}
\item First of all, as we motivated above, doing this 
on a single processor increases locality: if all points in a coloured block
(see the figure) fit in cache, we get reuse of the intermediate points.
\item Secondly, if we consider this as a scheme for distributed memory computation,
it reduces message traffic. Normally, for every update step the processors
need to exchange their boundary data. If we accept some redundant duplication
of work, we can now eliminate the data exchange for the intermediate levels.
The decrease in communication will typically outweigh the increase in work.
\end{itemize}

\Level 1 {Analysis}

Let's analyze the algorithm we have just sketched.  As in
equation~\eqref{eq:3p-average} we limit ourselves to a 1D set of
points and a function of three points. The parameters describing the
problem are these:
\begin{itemize}
\item $N$ is the number of points to be updated, and $M$~denotes the
  number of update steps. Thus, we perform $MN$ function evaluations.
\item $\alpha,\beta,\gamma$ are the usual parameters describing
  latency, transmission time of a single point, and time for an
  operation (here taken to be an $f$ evaluation).
\item $b$ is the number of steps we block together.
\end{itemize}
Each halo communication consists of $b$ points, and we do this $\sqrt
N/b$ many times.  The work performed consists of the $MN/p$ local
updates, plus the redundant work because of the halo. The latter term
consists of $b^2/2$ operations, performed both on the left and right
side of the processor domain.

Adding all these terms together, we find a cost of
\[ \frac Mb\alpha+M\beta+\left(\frac {MN}p+Mb\right)\gamma. \]
We observe that the overhead of $\alpha M/b+\gamma Mb$ is independent of~$p$,
Note that  the optimal value of~$b$ only depends on
  the architectural parameters $\alpha,\beta,\gamma$ but not on the
  problem parameters.

\Level 1 {Communication and work minimizing strategy}

\begin{figure}[ht]
\includegraphics[scale=.1]{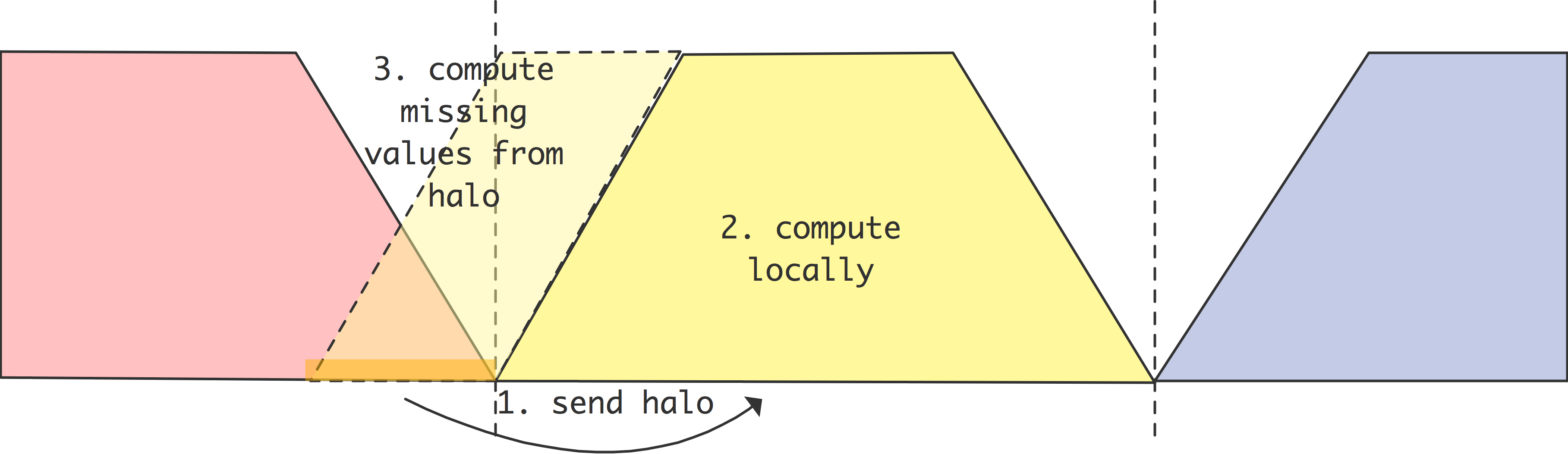}
\caption{Computation of blocks of grid points over multiple iterations}
\label{fig:grid-update-local}
\end{figure}

We can make this algorithm more efficient by overlapping the
communication and computation. As illustrated in
figure~\ref{fig:grid-update-local}, each processor start by
communicating its halo, and overlapping this communication with the
part of the communication that can be done locally. The values that
depend on the halo will then be computed last.

If the number of points per processor is large enough, the amount of
communication is low relative to the computation, and you could take
$b$ fairly large. However, these grid updates are mostly used in
iterative methods such as the \indexac{CG} method, and in that case
considerations of roundoff prevent you from taking $b$ too
large\cite{ChGe:sstep}.

\begin{figure}[ht]
\includegraphics[scale=.1]{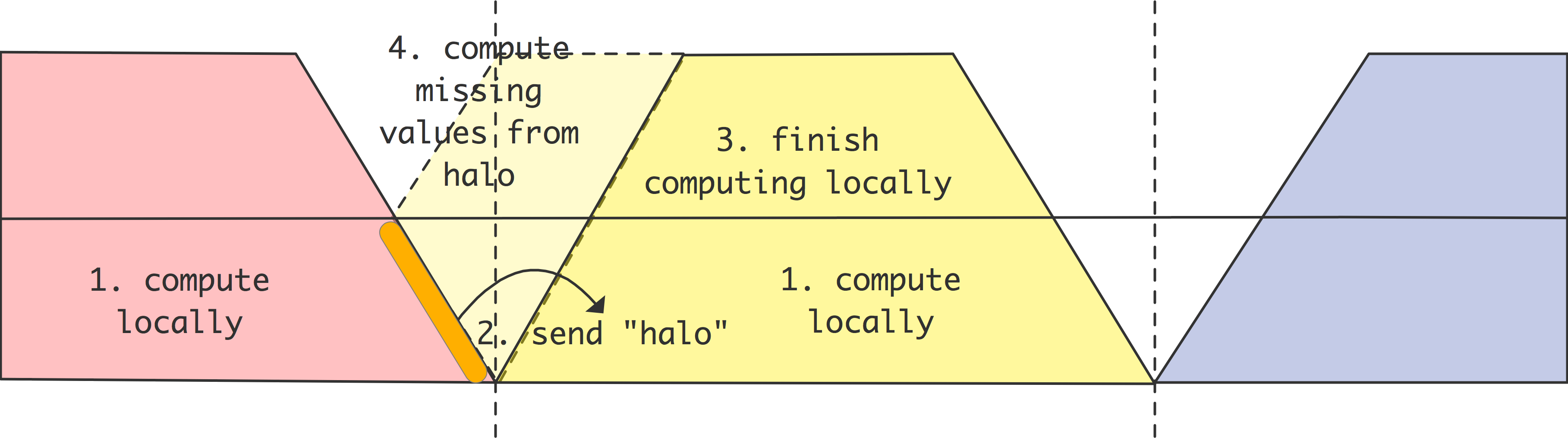}
\caption{Computation of blocks of grid points over multiple iterations}
\label{fig:grid-update-minimal}
\end{figure}

A further refinement of the above algorithm is possible.
Figure~\ref{fig:grid-update-minimal} illustrates that it is possible
to use a halo region that uses different points from different time steps.
This algorithm (see~\cite{Demmel2008IEEE:avoiding}) cuts down on the amount
of redundant computation. However, now the halo values that are communicated
first need to be computed, so this requires splitting the local communication
into two phases.

\section{A `communication avoiding' framework}

We now show how the above scheme, proposed for iterative methods, can
be applied to general task graphs. This means that we can have a
`communication avoiding compiler', that turns an arbitrary computation
into a communication avoiding one.


\label{sec:avoid-framework}

Above, we showed the
traditional strategy of communication a larger halo than would be
strictly
necessary~\cite{Douglas:caching-multigrid,Eijkhout:poly-smooth,OpJo:improved-ssor}.
With this, and some redundant computation, it is possible to remove
some synchronization
synchronization points from the computation.

However, this is not guaranteed to overlap communication and
computation; also, it is possible to avoid some of the redundant work.
We will now formalize this `communication avoiding'
strategy~\cite{Demmel2008IEEE:avoiding}.

\begin{figure}[ht]
  \includegraphics[scale=.4]{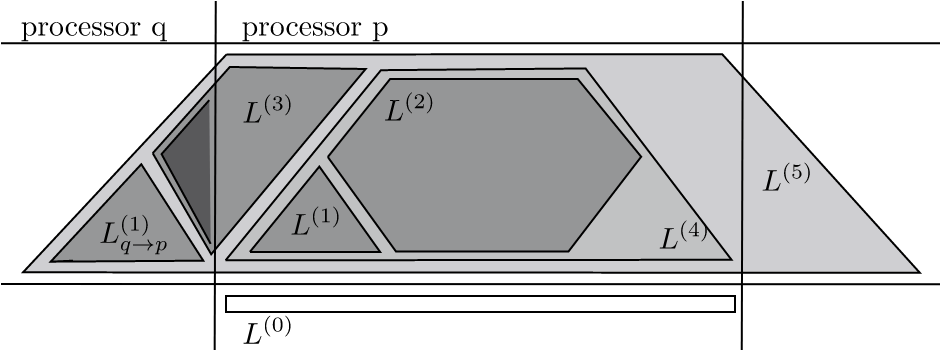}
  \caption{Subdivision of a local computation for minimizing communication and redundant computation.}
  \label{fig:avoid}
\end{figure}

We start with a distributed task graph
$\{ L_{p} \}_{p}$ with a predecessor relation
\[ t' \in \pred(t)\equiv \{\hbox{task $t'$ computes direct input data for
  task $t$}\}
\]
which holds on the global graph.

We now derive subsets
$L^{(1)}_{p},L^{(2)}_{p},L^{(3)}_{p}$
based on a formal latency-avoiding argument:
we will have
\[
    L_{p} \subsetneq L^{(1)}_{p} \cup L^{(2)}_{p} \cup L^{(3)}_{p}
\]
and any latency will be hidden by the computation of
$L^{(2)}_{p}$,
dependent of course on the size of the  original task graph.

We now derive a collection of subsets, not neceessarily disjoint, that
defines our latency tolerant computation.
These
concepts are illustrated in figure~\ref{fig:avoid}.

\heading{Subset 0: inherited from previous level}

We define
$L^{(0)}_{p}$
as the data that is available
on process~$p$
before any computation takes place:
\[
L^{(0)}_{p} \quad\hbox{contains initial conditions}
\]
This is either a true initial condition, or the final result of a
previous block step.

\heading{Subset 4: all locally computed tasks}

We define an auxiliary set
$L^{(4)}_{p}$
as the tasks in
$L_{p}$
that can be computed without
needed data from processors $q\not=p$. 
\[
L^{(4)}_{p} \equiv
\bigl\{ t\colon \pred(t)\in \{ L^{(0)}_{p} \cup L^{(4)}_{p} \} 
\bigr\}
\]
This is a subset of all the $L_p$ local tasks.

\heading{Subset 5: all predecessors of local tasks}
  
Next we define
$L^{(5)}_{p}$
as all tasks
that are computed anywhere to construct the local result
$L_{p}$:
\[
L^{(5)}_{p} \equiv
L_{p} \cup \pred(L_{p})
\]
This is a superset of the local tasks~$L_p$; it includes non-local
tasks (that is in $L_q$ for $q\not=p$) that would be communicated in a
naive computation.

\heading{Subset 1: locally computed tasks, to be used remotely}

We now define
$L^{(1)}_{p}$
as the locally
computed tasks on~$p$ that are needed for a $q\not=p$:
\[
L^{(1)}_{p}\equiv
L^{(4)}_{p} \cup \bigcup_{q\not=p} L^{(5)}_{q}
- L^{(0)}_{p}
\]
These are the tasks computed first. For each $q\not=p$, a subset of
these elements will be sent to process~$q$, in a communication step
that overlaps the computation of the next subset.

\heading{Subset 2: locally computed tasks, only used locally}

While elements of
$L^{(1)}_{p}$
are being sent, we can do a local
computation of the remainder of
$L^{(4)}_{p}$:
\[
L^{(2)}_{p} \equiv L^{(4)}_{p} - L^{(1)}_{p}
\]
These tasks use results from~$L_1$, but are otherwise entirely local,
since they are part of the local set~$L_4$.

\heading{Subset 3: halo elements and their successors}

The final part of the computation on~$p$ consist of those tasks that,
recursively, need results from other processors.

Having received remote elements
$L^{(1)}_{q\rightarrow p}$
from neighbouring
processors~$q\not=p$, we can construct the remaining elements
of~$L^{(5)}_{p}$ that are needed for~$L_{p}$:
\[
L^{(3)}_{p} \equiv
L^{(5)}_{p} - L^{(4)}_{p} - \bigcup_{q\not=p} L^{(1)}_{q}
\]

\begin{theorem}
  The splitting $L^{(1)},L^{(2)},L^{(3)}$ is well-formed and has overlap
  of communication $L^{(1)}\rightarrow L^{(3)}$ with the computation of $L^{(2)}$.
  Neither $L^{(1)}$ nor $L^{(2)}$ have synchronization points, so the whole algorithm
  has overlap.
\end{theorem}

However, note that $L^{(1)}\cup L^{(2)}\cup L^{(3)}$ is most
likely larger than~$L_{k}$,
corresponding to redundant calculation.


\begin{figure}[ht]
  \includegraphics[scale=.32]{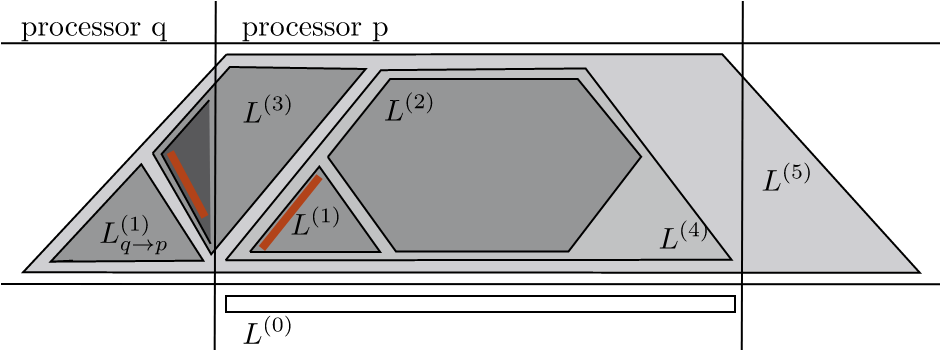}
  \caption{Communicated sets in the communication avoiding scheme}
  \label{fig:avoid-comm}
\end{figure}

In figure~\ref{fig:avoid-comm} we indicate in red the part of
$L^{(0)}$ that is sent, and the part of~$L^{(3)}$ that is received.

\section{Simulation}

\begin{figure}[ht]
  \includegraphics[scale=.15]{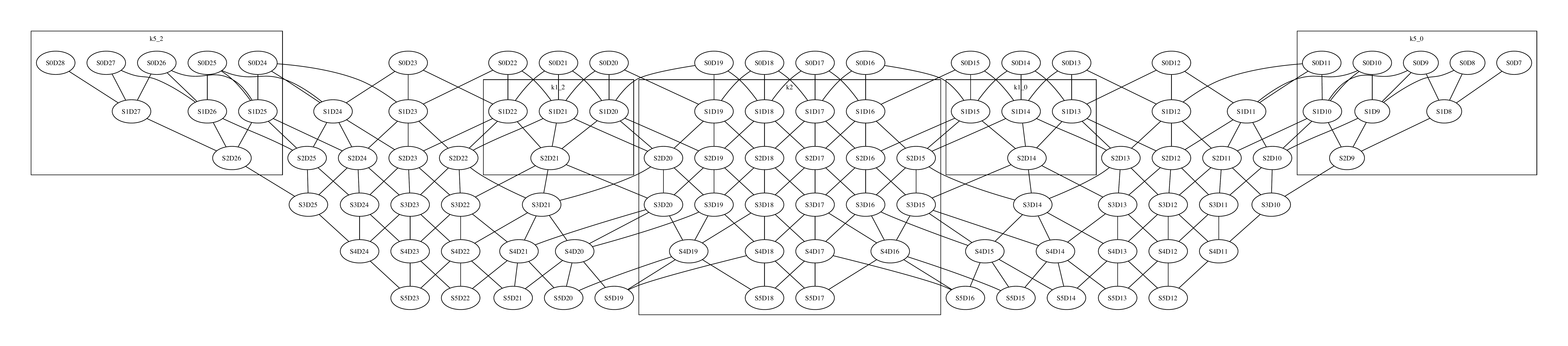}
  \caption{Depiction of $k_1,k_2,k_3$ sets for a processor doing a 1D
    heat equation.}
  \label{fig:1dk123}
\end{figure}

We have written a simple simulator\footnote{The code can be found in
  the code repository~\cite{IMPcode-repo} under {\tt pocs/avoid}.} that takes a task graph and
identifies the $k_1,k_2,k_3$ sets. In figure~\ref{fig:1dk123} we
illustrate these sets for a one-dimensional case, but we note that the
analysis works on arbitrary task graphs.

\begin{figure}[ht]
  \includegraphics[scale=.6]{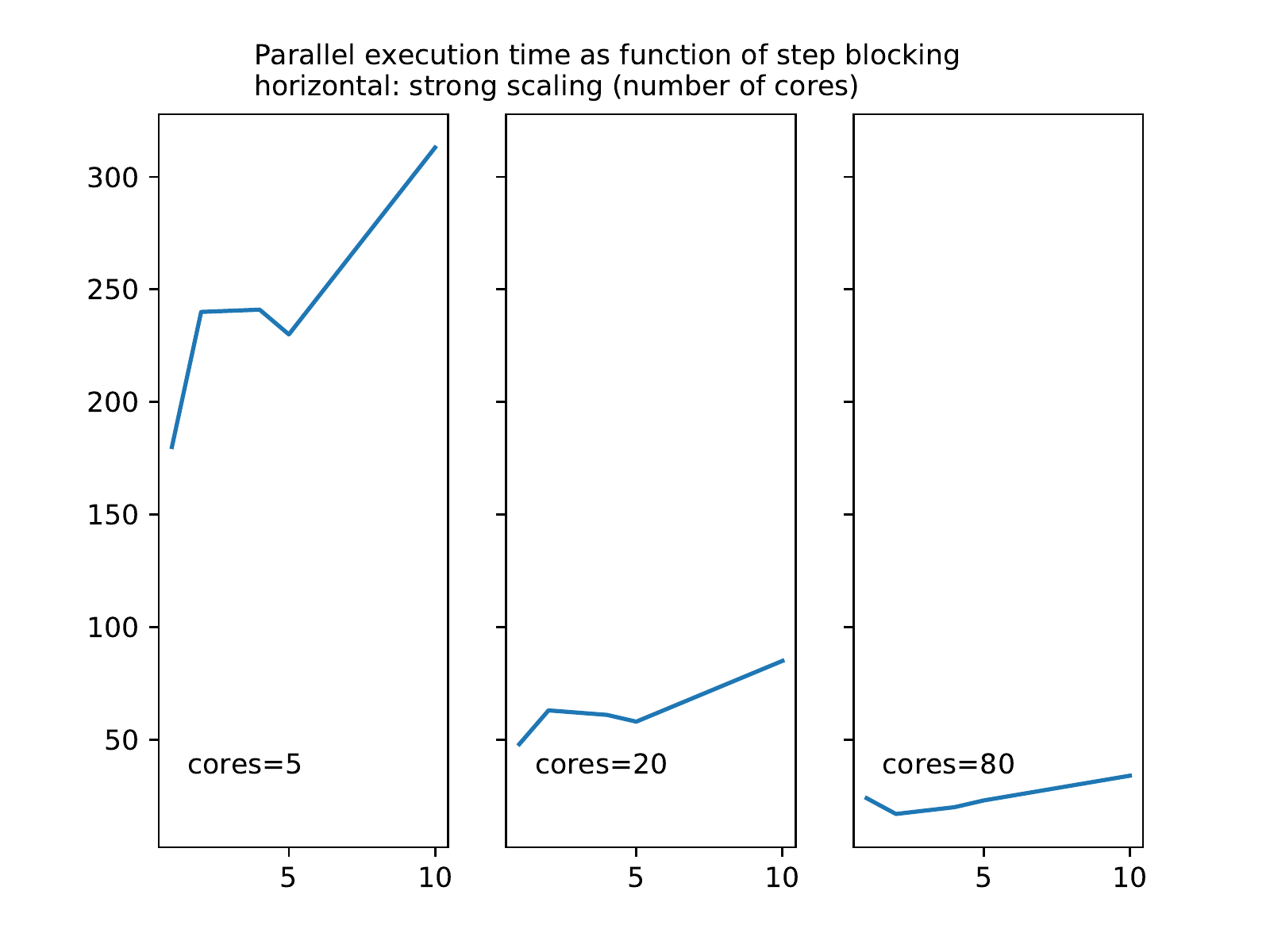}
  \caption{Runtime as a function of core count for moderate latency}
  \label{fig:lat1000}
\end{figure}
\begin{figure}[ht]
  \includegraphics[scale=.6]{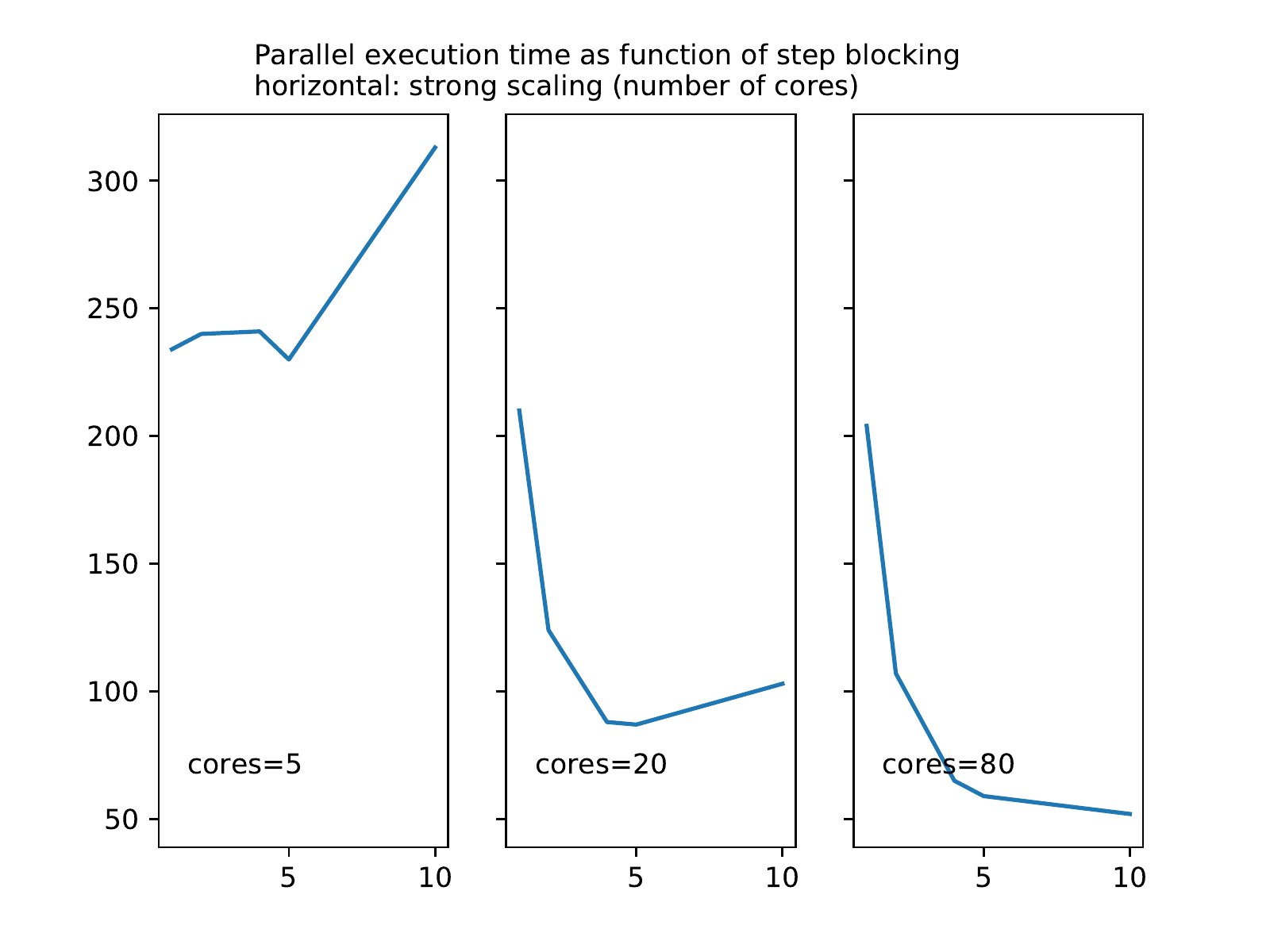}
  \caption{Runtime as a function of core count for high latency}
  \label{fig:lat10000}
\end{figure}

We then simulate parallel runtimes by evaluating runtimes for the
following scenario:
\begin{itemize}
\item We use a strong scaling scenario where we have a given problem
  size and partitioning into tasks, as well the number of MPI nodes;
\item We set the ratio of message latency to floating point operation
  as fixed;
\item We evaluate the runtime as a function of the number of threads
  available for the task graph on an MPI node.
\end{itemize}
The prediction is that, with non-negligible latency, blocking will
reduce the running time more than the extra computation time from the
extended halo. Also, this effect will be more pronounced as the core
count increases, since this reduces the no-node computation time.

First, in figure~\ref{fig:lat1000} we use a low latency; we see that
only for very high thread count is there any gain. In
figure~\ref{fig:lat10000} we use a higher latency, and we see that
even for moderate thread counts blocking effects latency hiding.

\section{Conclusion}

We have discussed the concepts and prior work in latency hiding and
communication avoiding. We have shown that the IMP framework can turn
an arbitrary computation graph into a communication avoiding one by
judicious partitioning of the task graph, and duplicating certain tasks.

\bibliography{vle,imp}

\begin{thebibliography}{10}

\bibitem{ChGe:sstep}
A.~Chronopoulos and C.W. Gear.
\newblock {$s$}-step iterative methods for symmetric linear systems.
\newblock {\em Journal of Computational and Applied Mathematics}, 25:153--168,
  1989.

\bibitem{DAzEijRo:ppscicomp}
E.F. D'Azevedo, V.L. Eijkhout, and C.H. Romine.
\newblock A matrix framework for conjugate gradient methods and some variants
  of cg with less synchronization overhead.
\newblock In {\em Proceedings of the Sixth SIAM Conference on Parallel
  Processing for Scientific Computing}, pages 644--646, Philadelphia, 1993.
  SIAM.

\bibitem{dehevo92:acta}
J.~Demmel, M.~Heath, and H.~{Van der Vorst}.
\newblock Parallel numerical linear algebra.
\newblock In {\em Acta Numerica 1993}. Cambridge University Press, Cambridge,
  1993.

\bibitem{Demmel2008IEEE:avoiding}
James Demmel, Mark Hoemmen, Marghoob Mohiyuddin, and Katherine Yelick.
\newblock Avoiding communication in sparse matrix computations.
\newblock In {\em IEEE International Parallel and Distributed Processing
  Symposium}, 2008.

\bibitem{Douglas:caching-multigrid}
C.~Douglas.
\newblock Caching in multigrid algorithms: problems in two dimensions.
\newblock {\em International Journal of Parallel, Emergent and Distributed
  Systems}, 9:195--?204, 1996.

\bibitem{Eijkhout:poly-smooth}
Victor Eijkhout.
\newblock Polynomial acceleration of optimised multi-grid smoothers; basic
  theory.
\newblock Technical Report UT-CS-02-477, Innovative Computing Lab, University
  of Tennessee Knoxville, August 2002.

\bibitem{IMPcode-repo}
Victor Eijkhout.
\newblock {IMP} code repository, 2014-7.
\newblock \url{https://bitbucket.org/VictorEijkhout/imp-demo}.

\bibitem{Eijkhout:mathematical2016arxiv}
Victor Eijkhout.
\newblock A mathematical formalization of data parallel operations.
\newblock {\em CoRR}, abs/1602.02409, 2016.
\newblock \url{http://arxiv.org/abs/1602.02409}.

\bibitem{Gropp:libraries}
Bill Gropp.
\newblock Update on libraries.
\newblock
  \url{http://jointlab-pc.ncsa.illinois.edu/events/workshop3/pdf/presentations/Gropp-Update-on-Libraries.pdf}.

\bibitem{GrSm:petsc}
W.~D. Gropp and B.~F. Smith.
\newblock Scalable, extensible, and portable numerical libraries.
\newblock In {\em Proceedings of the Scalable Parallel Libraries Conference,
  IEEE 1994}, pages 87--93.

\bibitem{Me:multicg}
Gerard Meurant.
\newblock Multitasking the conjugate gradient method on the {CRAY} {X-MP/48}.
\newblock {\em Parallel Computing}, 5:267--280, 1987.

\bibitem{OpJo:improved-ssor}
Thomas Oppe and Wayne~D. Joubert.
\newblock Improved {SSOR} and incomplete {Cholesky} solution of linear
  equations on shared memory and distributed memory parallel computers.
\newblock {\em Numerical Linear Algebra with Applications}, 1:287--311, 1994.

\bibitem{Sa:practicalKrylov}
Yousef Saad.
\newblock Practical use of some krylov subspace methods for solving indefinite
  and nonsymmetric linear systems.
\newblock {\em SIAM J. Sci. Stat. Comput.}, 5:203--228, 1984.

\end{thebibliography}
\bibliographystyle{plain}

\end{document}